\documentclass[12pt]{article}
\usepackage{sw20lart}
\usepackage{graphicx}
\usepackage{cite}

\input{tcilatex}
\begin{document}

\title{On Some Perturbation Approaches to Population Dynamics}
\author{Francisco M. Fern\'{a}ndez \thanks{%
e--mail: fernande@quimica.unlp.edu.ar} \\
%EndAName
INIFTA (UNLP, CCT La Plata--CONICET), \\
Divisi\'{o}n Qu\'{i}mica Te\'{o}rica,\\
Diag. 113 y 64 (S/N), Sucursal 4, Casilla de Correo 16,\\
1900 La Plata, Argentina}
\maketitle

\begin{abstract}
We show that the Adomian decomposition method, the time--series expansion,
the homotopy--perturbation method, and the variational--iteration method
completely fail to provide a reasonable description of the dynamics of the
simplest prey--predator system.
\end{abstract}

In is well known that a reasonable description of the dynamics of a
nonlinear system by means of a perturbation method is a difficult enterprize%
\cite{BO78}. Recently, several approaches have been proposed for the
treatment of the simplest model for the prey--predator interaction. They are
based on the Adomian descomposition method \cite{BM05}, the time--series
expansion\cite{BIK05}, the homotopy--perturbation method\cite{CHA07,RDGP07},
and the variational--iteration method\cite{YE08}. We have already shown that
the time--power series given by the implementation of the
homotopy--perturbation method proposed by Chowdhury et al\cite{CHA07}
completely fails to yield the main features of the population dynamics\cite
{F07}. In this short communication we briefly analyze the results of the
other proposals in terms of what one expects from an approach designed to
solve problems of population dynamics\cite{BO78}.

The chosen model is the prey--predator system\cite{BM05,BIK05,RDGP07,YE08}
\begin{eqnarray}
\frac{dx(t)}{dt} &=&x(t)[a-by(t)],\;a,b>0  \nonumber \\
\frac{dy(t)}{dt} &=&-y(t)[c-dx(t)],\;c,d>0  \label{eq:model}
\end{eqnarray}
where $x(t)$ and $y(t)$ are the populations of rabits and foxes,
respectively. This nonlinear system exhibits a saddle point at $%
(x_{s},y_{s})=(0,0)$ and a center at $(x_{s},y_{s})=(c/d,a/b)$\cite{BO78}.
Besides, the populations obey the following curve in the $x-y$ plane:
\begin{equation}
\ln \left( x^{c}y^{a}\right) -dx-by=\ln \left( x_{0}^{c}y_{0}^{a}\right)
-dx_{0}-by_{0}  \label{eq:curve}
\end{equation}
where $x_{0}$ and $y_{0}$ are the initial populations at time $t=0$.

The homotopy--perturbation method proposed by Rafei et al\cite{RDGP07} leads
to time--series expansions of the form
\begin{eqnarray}
x(t) &=&x_{0}+x_{0}(a-by_{0})t+\ldots  \nonumber \\
y(t) &=&y_{0}+y_{0}(dx_{0}-c)t+\ldots  \label{eq:t-series}
\end{eqnarray}
that are exactly the same as those otained earlier by Biazar et al\cite
{BIK05}. Obviously, they are suitable about the point $(x_{0},y_{0})$ and
will not predict the main features of the population dynamics (revealed at a
much greater time span) which is what really matters in this field\cite{BO78}%
. The variational--iteration method\cite{YE08} yields more complicated
expressions but the authors state that they agree with the time--power
series mentioned above.

\begin{figure}[]
\begin{center}
\includegraphics[width=9cm]{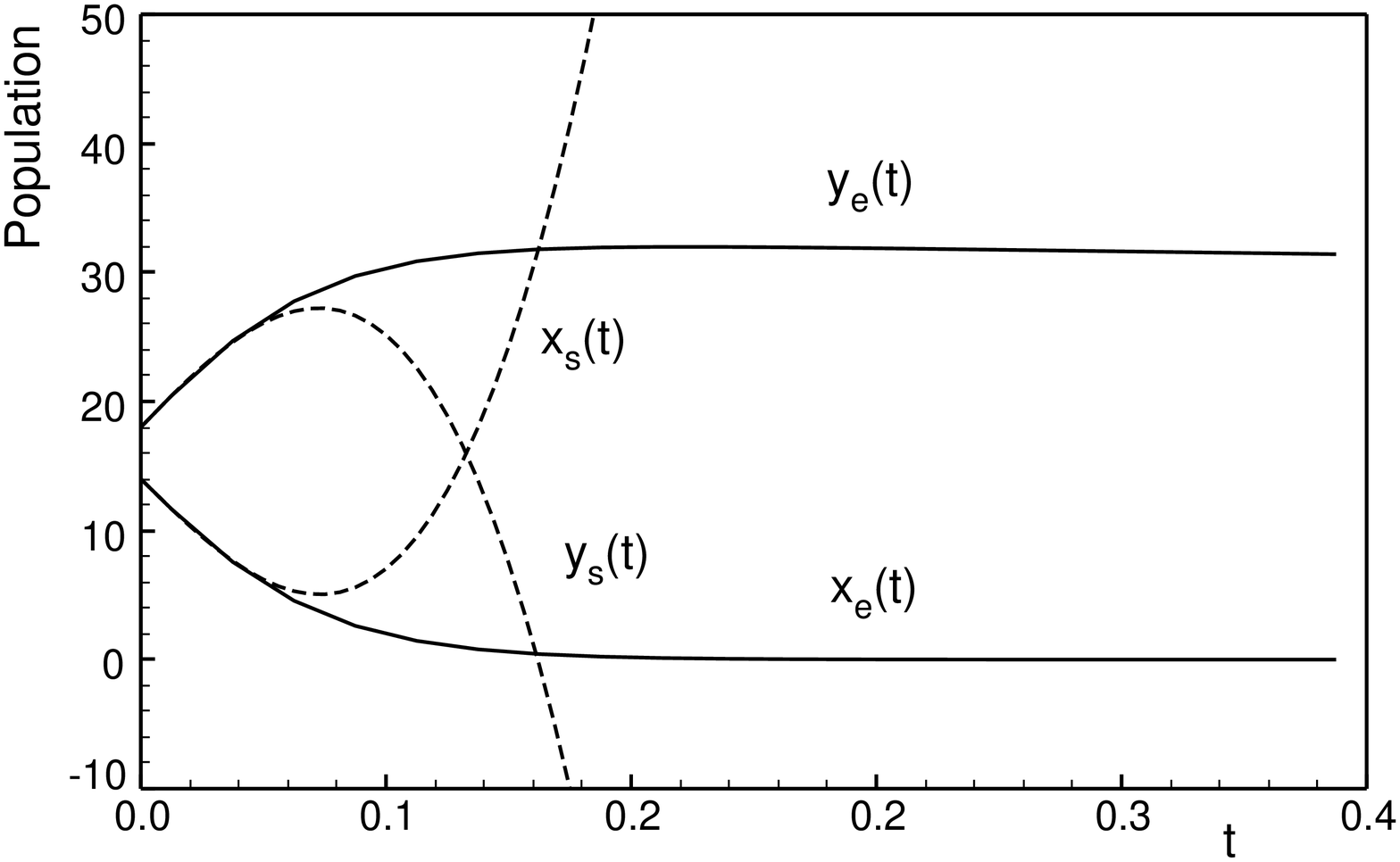}
\end{center}
\caption{Exact (solid, e) and series (dashed, s) populations for Case I}
\label{Fig:RDGP1}
\end{figure}

Fig.~\ref{Fig:RDGP1} shows the populations for Case I ($a=b=d=1$, $c=0.1$, $%
x_{0}=14$, $y_{0}=18$) in a time span larger than that considered in the
earlier studies already mentioned above\cite{BM05,BIK05,RDGP07,YE08}. We
clearly appreciate that all those approaches predict a wrong behaviour of
the populations. Besides, the time span in Fig.~\ref{Fig:RDGP1} is still not
large enough to cover the whole portrait of the system evolution.

\begin{figure}[]
\begin{center}
\includegraphics[width=9cm]{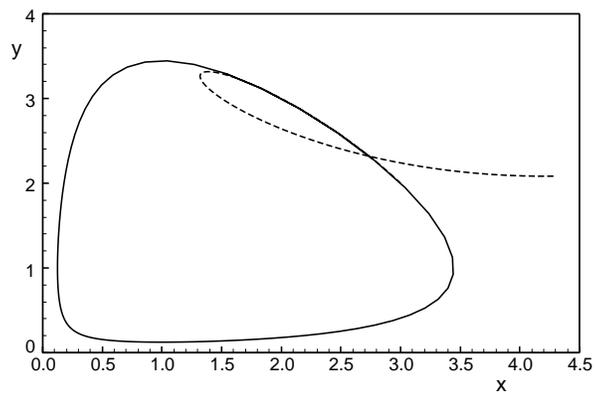}
\end{center}
\caption{Exact (solid) and approximate (dashed) populations for Case V in
the $x-y$ plane}
\label{Fig:RDGP2}
\end{figure}

Case V ($a=b=c=d=1$, $x_{0}=3$, $y_{0}=2$) is much more
interesting\cite{BIK05,RDGP07}.
Fig.~\ref{Fig:RDGP2} shows the results in the $x-y$ plane.
The time series\cite{BIK05,RDGP07}
predict a completely wrong behaviour and the resulting curve cross itself
which can never happen as it is well known\cite{BO78}.

Summarizing: present analysis clearly shows that the Adomian descomposition
method\cite{BM05}, the
straightforward time--series expansion\cite{BIK05}, the homotopy--perturbation
method\cite{RDGP07,CHA07}, and the
variational--iteration method\cite{YE08} are completely useless for a
reasonable prediction of the evolution of even the simplest prey--predator
systems.

\end{document}